\documentclass[prl,twocolumn,floatfix,superscriptaddress,a4paper,nofootinbib]{revtex4}
\usepackage{amsmath,bm,graphicx}
\usepackage{amssymb}
\usepackage{amsfonts}
\usepackage{epstopdf}
\usepackage{mathrsfs}
\usepackage{color}
\usepackage{latexsym}
\usepackage{hyperref}
\usepackage{subfigure}
\usepackage{bm}
\usepackage{natbib}

\newcommand{\Pe}{\mathrm{Pe}}

\begin{document}

\title{Splitting probabilities for dynamics in corrugated channels: passive VS active Brownian motion}

\author{Paolo Malgaretti}
\email[Corresponding Author: ]{p.malgaretti@fz-juelich.de }
\affiliation{Helmholtz Institute Erlangen-N\"urnberg for Renewable Energy (IEK-11), Forschungszentrum J\"ulich, Cauerstr. 1,
91058 Erlangen, Germany}

\author{Tatiana Nizkaia}
\affiliation{Helmholtz Institute Erlangen-N\"urnberg for Renewable Energy (IEK-11), Forschungszentrum J\"ulich, Cauerstr. 1,
91058 Erlangen, Germany}

\author{Gleb Oshanin}
\affiliation{Sorbonne Universit\'e, CNRS, Laboratoire de Physique Th\'eorique de la Mati\`ere Condens\'ee (UMR CNRS 7600), 4 place Jussieu, 75252 Paris Cedex 05, France}

\begin{abstract}
In many practically important problems which rely on particles'  transport
in realistic corrugated channels, one is interested to know  
the probability that either of the extremities, (e.g., the one containing a chemically active site, or connected to a broader
channel), 
is reached before the other one. In mathematical literature, the latter are called the "splitting" probabilities (SPs). 
Here, within the Fick-Jacobs approach, we study analytically the SPs as functions of system's parameters for dynamics in three-dimensional corrugated channels,  confronting standard diffusion and active Brownian motion. Our analysis reveals some similarities in the behavior and also some markedly different features, which can be seen as fingerprints
of the activity of particles.
\end{abstract}
\maketitle

Transport of particles in narrow
corrugated channels is an important area of research 
which has attracted a great deal of attention 
within the recent several decades (see e.g., Ref. \cite{a} for a review). 
In part, such an interest is due 
to the relevance to various realistic physical, biophysical and chemical systems, as well applications in nanotechnology and nanomedicine, e.g.,  for manufacturing of artificial molecular nanofilters. To name just a few 
examples, we mention transport in porins \cite{porins1,porins2}, in nuclear pores \cite{cell1,cell2,cell3}, in microtubules \cite{tubes} and dendritic spines \cite{spine}, transport of microswimmers in capillaries \cite{cap1,cap2}, translocation of polymers in pores \cite{transloc0,transloc1,transloc2} and their sequencing in nanopore-based devices \cite{seq}, as well as 
in microfluidics \cite{mf1,mf2}. 

The problem of random transport in corrugated channels is clearly
also a  challenge for the theoretical analysis - it is too complicated to be solved analytically in full detail and 
one therefore seeks approximate approaches that are justified in particular limits. 
Most of the available
analytical descriptions rely on the so-called Fick-Jacobs
approach \cite{merkel,zw} and its subsequent generalizations (see, e.g., \cite{rubi,Malgaretti2013,dean,dean2,dean3}). In essence,
this approach amounts to a
reduction of the original
multidimensional problem to 
a one-dimensional
diffusion in presence of some potential, which mimics in an effective way 
a spatial variation of the confining boundaries. 
In some cases, this approximation is
physically meaningful and provides an
insight into the behavior of important characteristic properties, 
e.g.,  currents across the channel,
the mean first-passage times to some positions and
 quantifying fluctuations of the first-passage times \cite{transloc3}. In other systems, in which, e.g.,
diffusion in the direction perpendicular to the main axis of the channel is important   \cite{val}, other approaches are to be developed.
\begin{figure}
\centering
\includegraphics[width=0.75\columnwidth]{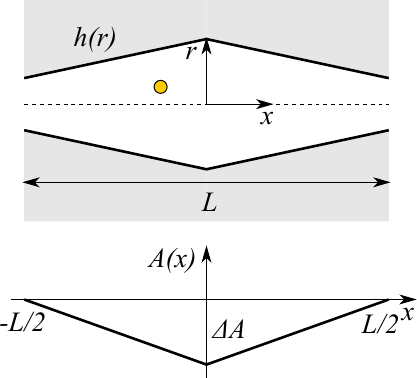}
\caption{Top: A colloidal particle in a simple varying-section channel with fore-aft symmetry. 
Bottom: the effective piece-wise linear potential $A(x)$ within the Fick-Jacobs approach and the corresponding overall barrier $\Delta A$.}
\label{fig:scheme_pot}
\end{figure}

In many important situations one is interested in understanding the behavior of the
properties which characterize a kind of a "broken symmetry" in otherwise symmetric 
dynamics : in particular,  of the probability that a particle injected at some position within the channel reaches first its prescribed 
extremity without having ever reached the opposite one. This particle can be a tracer within a channel that is attached to a broader pathway to which all the channels are connected, or it can be a chemically active molecule which needs to react with a target site placed at either of the extremities. In mathematical and physical literature (see, e.g., \cite{red,ben})
such probabilities - the so-called splitting probabilities - have been analyzed in details in various settings, with and without an external potential (see, e.g. \cite{osh}), providing an important
 complementary insight into the dynamical behavior.

In the present paper, we study analytically the behavior of  splitting probabilities (SPs) as functions of system's parameters  for transport in narrow corrugated channels,  
in terms of a suitably generalized Fick-Jacobs approach. In regard to the dynamics, we confront two different transport mechanisms - standard Brownian motion and active Brownian motion, 
capitalizing for the latter case on the theoretical framework developed in recent \cite{kal,kal2,kal3,kal4,kal5,kal6,kal7}.  For 
passive diffusion we obtain exact expressions for the SPs for channels of an arbitrary periodic shape. For the active case for which the dynamic equations  have a much more cumbersome form   \cite{kal,kal2,kal3,kal4,kal5,kal6,kal7}, we resort to a numerical analysis. 
  Our theoretical  findings demonstrate that the SPs are quite sensitive to both the geometry of the channel and 
  the activity of the particles. In particular, for active particles the SPs exhibit a spectacular non-monotonous dependence on the amplitude of the corrugation of the channel  when the  magnitude of the entropic force emerging due to a confinement becomes comparable to the propulsive force. This effect is absent for a passive Brownian motion.

\textbf{Passive particles.}
Consider a particle that starts at  $\mathbf{r}_0$ and undergoes a passive Brownian motion within an axially-symmetric three-dimensional channel 
with impermeable periodically-corrugated boundaries. It is convenient to use the cylindric coordinates $(r,x)$, where the $x$-axis coincides with the main axis of the channel, while $r$ is the radial coordinate. 
A local  
 thickness of the channel at point $x$ is defined by $h(x)$ and hence, $r \leq h(x)$. In view of the symmetry, the particle's position probability
 density function $\rho(\mathbf{r},t)$ and therefore all other properties derived from it are independent of the polar angle.
 We focus on the SPs - the probabilities that the particle first reaches either of the extremities $x = \pm L/2$  of the channel (see Fig. \ref{fig:scheme_pot}) without ever hitting the other one.

We first write down the advection-diffusion equation that governs the time evolution of the particle's position probability
 density function $\rho(\mathbf{r},t)$ : 
\begin{align}
\dot{\rho}(\mathbf{r},t)=\nabla\cdot \left[D\nabla\rho(\mathbf{r},t)+D\beta \rho(\mathbf{r},t) \nabla W(\mathbf{r})  \right]
\label{eq:adv-diff}
\end{align}
where $\mathbf{r}$ is the position of a particle, 
$D$ is the diffusion coefficient, $\beta^{-1}=k_BT$ is the inverse thermal energy, $k_B$ is the Boltzmann constant, $T$ - the absolute temperature and $W(\mathbf{r})$ is the particle-wall interaction potential, 
\begin{align}
W(\mathbf{r})=\begin{cases}
\phi(\mathbf{r}) , & r < h(x) , \\
\infty ,  & \text{otherwise.}
\end{cases}
\end{align}
If a local thickness of the channel is a slowly varying function of the $x$-coordinate, such that $\partial_x h(x)\ll 1$, it is possible to write down
the probability density function in the following approximate factorized form (see, e.g., ~\cite{Malgaretti2016}), 
\begin{align}
\label{rho}
\rho(\mathbf{r},t)=p(x,t)\dfrac{e^{-\beta W(\mathbf{r})}}{e^{-\beta A(x)}} \,,
\end{align}
where 
\begin{align}
A(x)=-\dfrac{1}{\beta}\ln\left[\frac{1}{\pi h_0^2}\int_{0}^{\infty} e^{-\beta W(x,r)} r dr\right]
\label{eq:def-A}
\end{align}
is the local free energy and $h_0$ is the mean cross-section of the channel. 
Upon integrating over the radial coordinate, we cast Eq. \eqref{rho} into the form:
\begin{align}
\dot{p}(x,t)=\partial_x\left[D\partial_x p(x,t)+D\beta p(x,t)\partial_x A(x) \right]
\label{eq:f-j}
\end{align}
Such a reduction of the original three-dimensional problem to 
a one-dimensional diffusion in presence of an effective potential (which, in fact, is the local free energy defined in Eq.~\eqref{eq:def-A}) is called the Fick-Jacobs approach~\cite{Zwanzig,Reguera2001,Malgaretti2013} and its range of applicability is well-understood~\cite{Reguera2006,berezhkovskii2007diffusion,Burada2007,dagdug2015,Kalinay2005,Kalinay2005_2,Kalinay2006,Martens2011,Dagdug2012_2,Dagdug2015_2}.
 This approach has provided an insight into the behavior of quite diverse confined systems, including colloidal particles \cite{Hanggi2011,dagdug2015}, flow of charged fluids~\cite{Hanggi2013,Malgaretti2014,Chinappi2018,Malgaretti2019JCP,Kalinay2020}, of polymers~\cite{Bianco2016,transloc3,Carusela2021}, of rigid rods~\cite{Malgaretti2021},  systems with chemical reactions~\cite{Ledesma-Duran2016}, and pattern-forming ones~\cite{Chacon-Acosta2020}.

We quantify next the SP $E_+$ - the probability that the particle first reaches $x=L/2$ without ever touching $x=-L/2$.  
This SP has the form  (see e.g. \cite{osh}) 
\begin{align}
E_+=\frac{\tau_-}{\tau_+ +\tau_-}\,,\quad E_-=\frac{\tau_+}{\tau_+ +\tau_-}\,,
\label{eq:hitting}
\end{align}
where $\tau\pm=\rho_0 L/|J_\pm|$ and $|J\pm|$ are the magnitudes of the steady-state currents from $x_0$ to the extremities $x=L/2$ and $x = - L/2$, respectively.  Note that the SP $E_-$ (i.e., the probability that the particle first reaches $x=- L/2$ without ever touching $x=L/2$) is simply defined by $E_-
= 1 - E_+$.

Solving Eq.~\eqref{eq:f-j}, we determine the steady-state currents $J\pm$ (see appendix) and hence, the  functions
$\tau_{\pm}$ to get
\begin{align}
\label{zz}
\tau_-=\tau_0\int_{-L/2}^{x_0} e^{\beta A(x)}dx\,,\quad \tau_+=\tau_0\int_{x_0}^{L/2} e^{\beta A(x)}dx \,, 
\end{align}
with $\tau_0=(L/D) e^{-\beta A(L)}$. Expressions \eqref{zz} totally define
 the SPs $E_{\pm}$. 
They are fairly general and hold for arbitrary $A(x)$, i.e., confining boundaries of arbitrary (sufficiently smooth) shapes. In the trivial case $A(x) \equiv 0$, we find from Eq. \eqref{zz} that the functions $\tau_{\mp} = L \left(L/2 \pm x_0 \right)/D$ and hence, recover the well-known result \cite{red}
\begin{align}
\label{z3}
E_+=\frac{1}{2} + \frac{x_0}{L} \,, \quad  E_-=\frac{1}{2} - \frac{x_0}{L} \,, \quad  -\frac{L}{2} \leq x_0 \leq \frac{L}{2}
\end{align} 
We will use Eq. \eqref{z3} in what follows as a point of reference
 - all departures from a simple linear behavior are indicative of the effects of the confining boundaries.

In order to get an idea of the dependence of the SPs on the overall barrier $\Delta A$ (see Fig. \ref{fig:scheme_pot}),
consider a simple  form of the 
free energy :
\begin{align}
\beta A(x) = \begin{cases}
\beta \frac{2\Delta A}{L}\left(x+\frac{L}{2}\right) , & -\frac{L}{2}<x\leq 0 ,\\
\beta \frac{2\Delta A}{L}\left(x-\frac{L}{2}\right) , & 0<x<\frac{L}{2} 
\end{cases}
\end{align}
We note parenthetically that such a simple piece-wise linear form has provided qualitatively reliable predictions on the behavior of the mean first-passage times through a finite channel in case of ions in a charged confinement~\cite{Malgaretti2016}.

\begin{figure}
\centering
\includegraphics[scale=0.55]{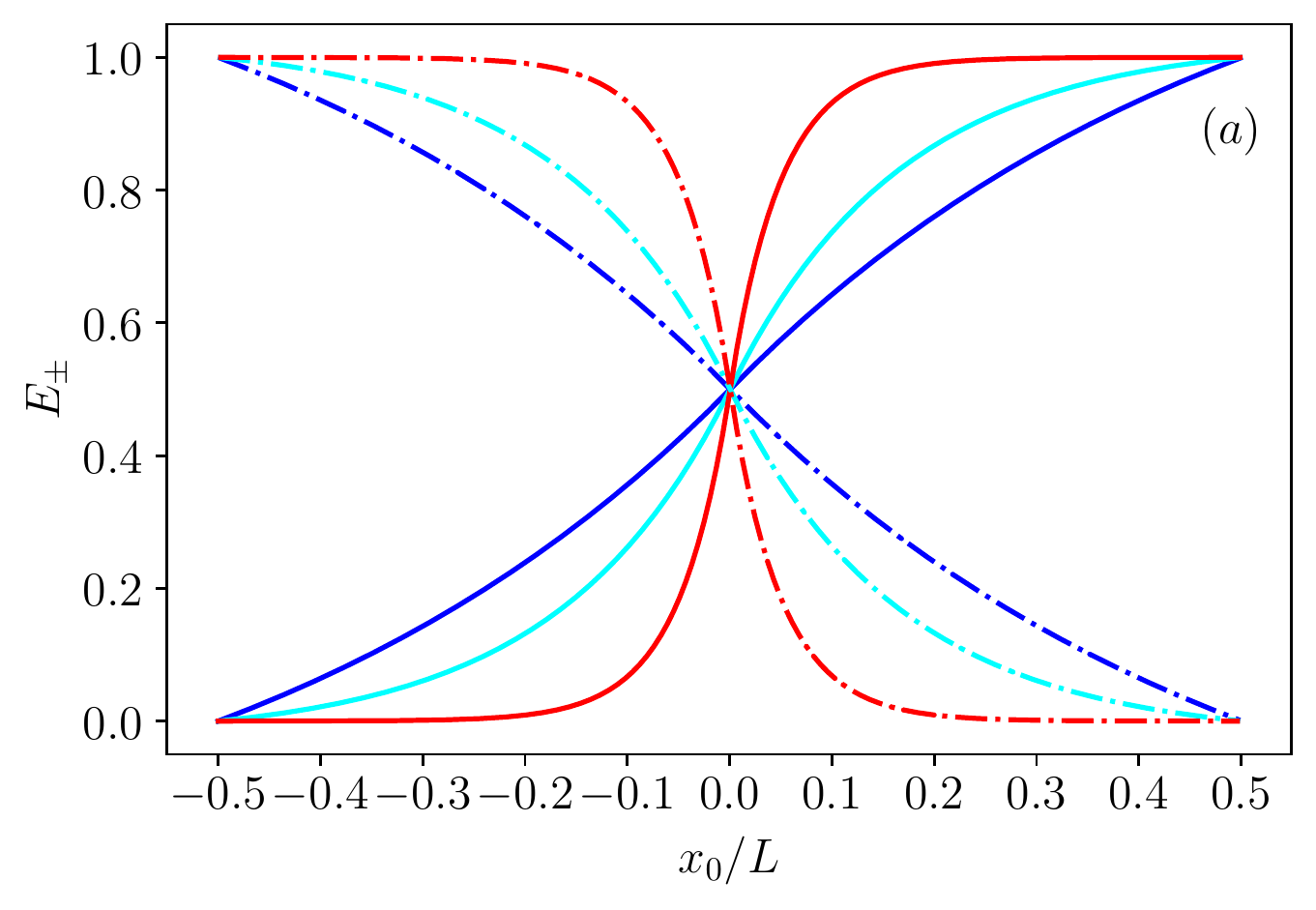}
\includegraphics[scale=0.55]{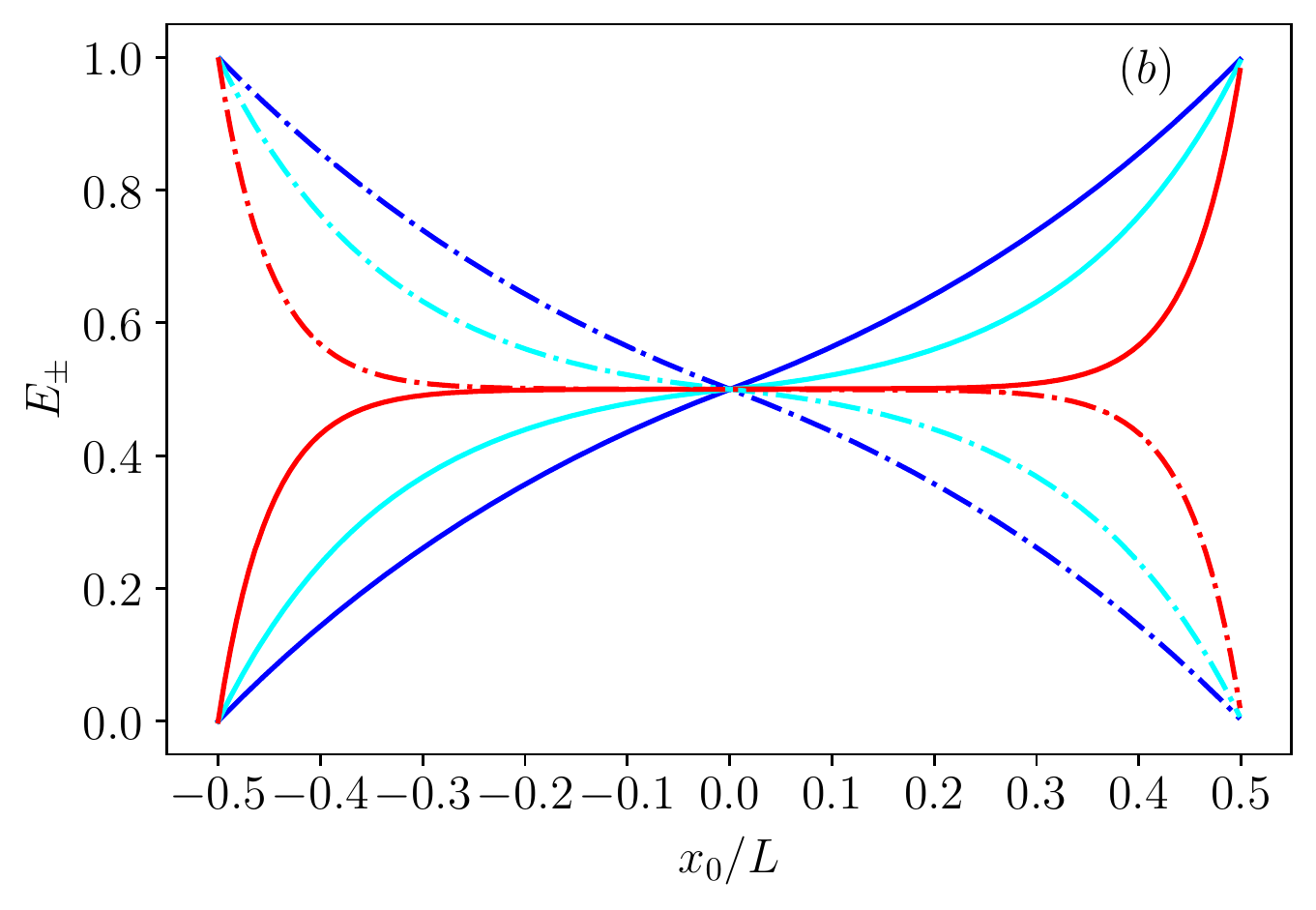}
\caption{$E_\pm$ as functions of the staring point $x_0/L$. Panel (a): $\beta \Delta A=1.0$ (blue), $3.0$ (cyan) and $10.0$ (red). Panel (b): $\beta \Delta A=-1.0$ (blue), $-3.0$ (cyan) and $-10.0$ (red). Solid lines stand for $E_+$ whereas dashed lines for $E_-$.}
\label{fig:x0}
\end{figure}
Fig.\ref{fig:x0} displays the SPs $E_\pm$ as functions of the starting point $x_0$ in the case of a fore-aft symmetric channel. We observe that for moderate values of the barrier, $\beta \Delta A= \pm 1$, i.e., for a mild  corrugation 
the SPs exhibit an almost linear dependence on $x_0$ (see blue curves in Fig. \ref{fig:x0}).   
For larger values of $\beta |\Delta A|$, the corrugation of the channel starts to play a major role and entails an essential departure from the linear dependence. We see that upon an increase of $\beta \Delta A$ to larger positive values, $E_+$ and $E_-$ attain an $S$-shaped form which becomes progressively more steep in the vicinity of $x = 0$ the larger $\beta \Delta A$ is. Recall that
for $\beta \Delta A>0$ the potential has a maximum at $x=0$ meaning that the channel has a bottleneck at this position. In consequence, when $x_0$ even only slightly exceeds $0$, it becomes much more probable for a particle to reach the right extremity because the bottleneck does not permit to reach the left one. Conversely, for  
 $\beta \Delta A<0$ the channel is widest at $x=0$. In this case, 
 if a particle starts in a broad part of the channel, it
 first diffuses there for a long time effectively "forgetting" about its actual starting point. Moreover, since in this case the channel narrows close to the extremities, there emerge effective entropic barriers (see, e.g., \cite{denis}) which the particle has to overcome 
 in order to reach any of the extremities. As a consequence, a passage to the extremity may necessitate repeated unsuccessful attempts to overpass the entropic barrier, which attempts are   interspersed with the excursions in the broad part of the channel.
 A combined effect of these two factors results in a very weak  dependence  of $E_{\pm}$ on $x_0$, which behavior we indeed observe in the panel (b) of  Fig. \ref{fig:x0}.

\begin{figure}
\centering
\includegraphics[scale=0.55]{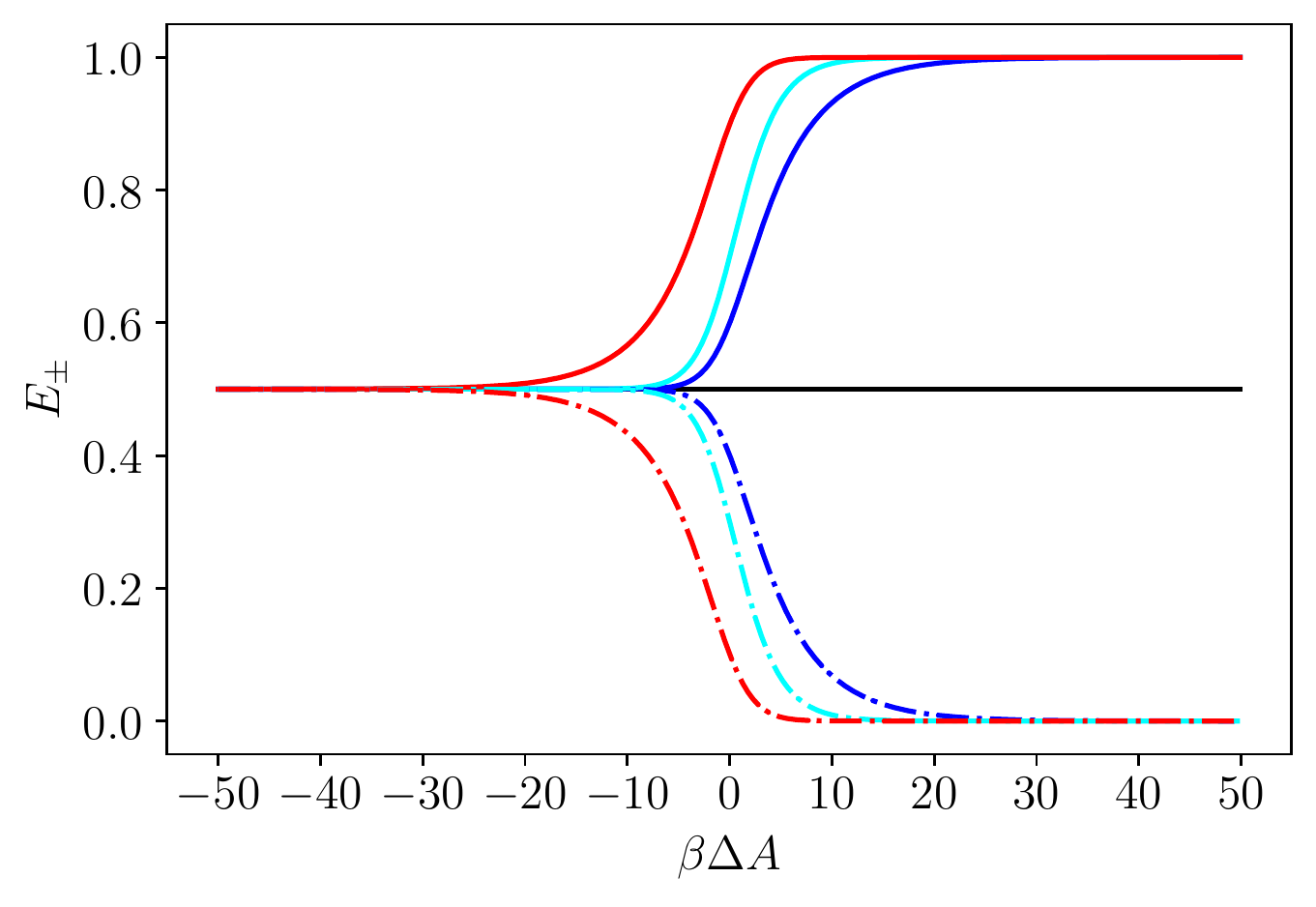}
\caption{$E_\pm$ as function of the entropic barrier $\beta \Delta A$ and $x_0/L=0.1$ (blue), $ 0.2$ (cyan) and $0.4$ (red).  Solid lines stand for $E_+$ whereas dashed lines -  for $E_-$.}
\label{fig:DA}
\end{figure}

Next, we examine the  dependence of $E_\pm$ on the magnitude of the barrier $\beta \Delta A$ for a few values of the starting position $x_0$. Figure~\ref{fig:DA} shows that for $\beta \Delta A \gg 1$ (i.e., for a strong entropic repulsion from the bottleneck at $x=0$) the SPs are either (almost) equal to zero or to unity, meaning that the particle most likely reaches first  the closest extremity and never gets to the  opposite one.  
In contrast, for $\beta \Delta A \ll -1$ (i.e., for an entropic repulsion from the extremities) the barrier to overcome becomes very high and a particle   has to undertake 
many  attempts to cross the  barrier before it actually does it. As a  consequence, for large negative $\beta \Delta A$ the SPs  $E_\pm \simeq 0.5$. 

It is important to emphasize  that  within the Fick-Jacobs approach, many characteristic properties of a particle diffusing in a channel, such as charge, elastic moduli, deformability, are effectively encoded in the free energy barrier $\Delta A$\cite{Malgaretti2016_entropy,Bianco2016,Malgaretti2019JCP,transloc3}. For example, for uncharged particles which are much smaller than the channel bottleneck (i.e., point-like particles) we have $\beta \Delta A\lesssim 3$ where $\beta \Delta A = 3$ implies that  the maximal cross-section of the channel is $\sim 30$ times the radius  of the bottleneck. In contrast, for charged ions it is feasible to have  $\beta \Delta A \simeq 10$ when electrostatic potential at the walls is $\beta e \zeta \simeq 10$~\cite{Malgaretti2019JCP}. Finally, for deformable objects, like polymers, one may have a very large  effective barrier $\beta \Delta A \simeq 100$~\cite{Bianco2016,Orlandini2017,transloc3}.

\textbf{Active particles.}
The case of particles that propel themselves through the channels, e.g., of "active" colloids, 
is most challenging, because
a local violations of the equilibrium may lead to quite a different  scenario as compared to the case of a passive Brownian motion. To set-up the scene, consider first a simple situation in which the non-interacting particles 
move with a constant velocity either to the left or to the right in a 
 one-dimensional system and interchange the sign of the velocity at random, at a constant rate $\alpha$.
In such a model  the time evolution of the densities $\rho_\downarrow(x,t)$  and $\rho_\uparrow(x,t)$ of active particles moving to the left or to the right, respectively, is described by ~\cite{Malgaretti2021}: 
\begin{eqnarray*}
\dot{\rho}_{\uparrow} & = & D\partial_{x}\left[\partial_{x}\rho_{\uparrow}+\beta\rho_{\uparrow}\partial_{x}A(x)+\beta\rho_{\uparrow}F_{act}\right]-\alpha\rho_{\uparrow}+\alpha\rho_{\downarrow}\label{eq:smol_1}\\
\dot{\rho}_{\downarrow} & = & D\partial_{x}\left[\partial_{x}\rho_{\downarrow}+\beta\rho_{\downarrow}\partial_{x}A(x)-\beta\rho_{\downarrow}F_{act}\right]+\alpha\rho_{\uparrow}-\alpha\rho_{\downarrow}\label{eq:smol_2}
\end{eqnarray*}
where $F_{act}$ is the propulsive force. These equations are to be solved subject to the 
 boundary conditions imposed at the extremities : 
$\rho_{\uparrow,\downarrow}(x=x_0)  =  \rho_{0}/2$ and 
$\rho_{\uparrow,\downarrow}(x=\pm\frac{L}{2})  =  0.$
A straightforward analysis~\cite{Malgaretti2021} shows that the dynamical behavior  
is characterized by two dimensionless parameters: the P\'eclet number
$\text{Pe}=\beta F_{act} R$ and the reduced hopping rate $\Gamma= \alpha R^2/D$. In particular, for a Janus swimmer~\cite{Malgaretti2021} the hopping between two states stems from a rotational diffusion of the particle and  $\Gamma_{*}=3/4$. For other types of swimmers (e.g., bacteria), the hopping rates can be much smaller.
Rewriting the equations in dimensionless form (but keeping the same notations), and setting the length scale to $L$, 
we have that the particle probability densities $\rho=\rho_0^{-1}(\rho_\uparrow +\rho_\downarrow)$ and $\phi=\rho_0^{-1}(\rho_\uparrow - \rho_\downarrow)$ obey, in the steady state,
\begin{equation}
\begin{split}
0 &= \partial_x[\partial_{x}\rho+\beta \Delta A\rho\partial_{x}a(x)+ \dfrac{\Pe L}{R}\phi]  ,\\
0 &=\partial_{x}\left[\partial_{x}\phi+\beta \Delta A\phi\partial_{x}a(x)+\dfrac{\Pe L}{R}\rho\right] -2\dfrac{L^2}{R^2}\Gamma\phi ,
\label{eq:rho_phi}
\end{split}
\end{equation}
while  the boundary conditions take the form
\begin{equation}
\rho(x_0)=\rho_0,  \rho(\pm 1/2)=0,\,
\phi(x_0)=0,  \phi(\pm 1/2)=0.
\end{equation}
Here $a(x)$ is a piecewise-linear function such that $a(0)=1$ and $a(\pm 1/2)=0$.

Note that the reduced channel length $L/R$ appears in the equations only in a combination with $\Pe$ and $\Gamma$. This means that the SPs for various channel lengths can be obtained by taking the solution at fixed $L/R$ and changing $\text{Pe}$ and $\Gamma$ accordingly. In the following we use $L/R$=10.

\begin{figure}[h]
\centering
\includegraphics[scale=0.55]{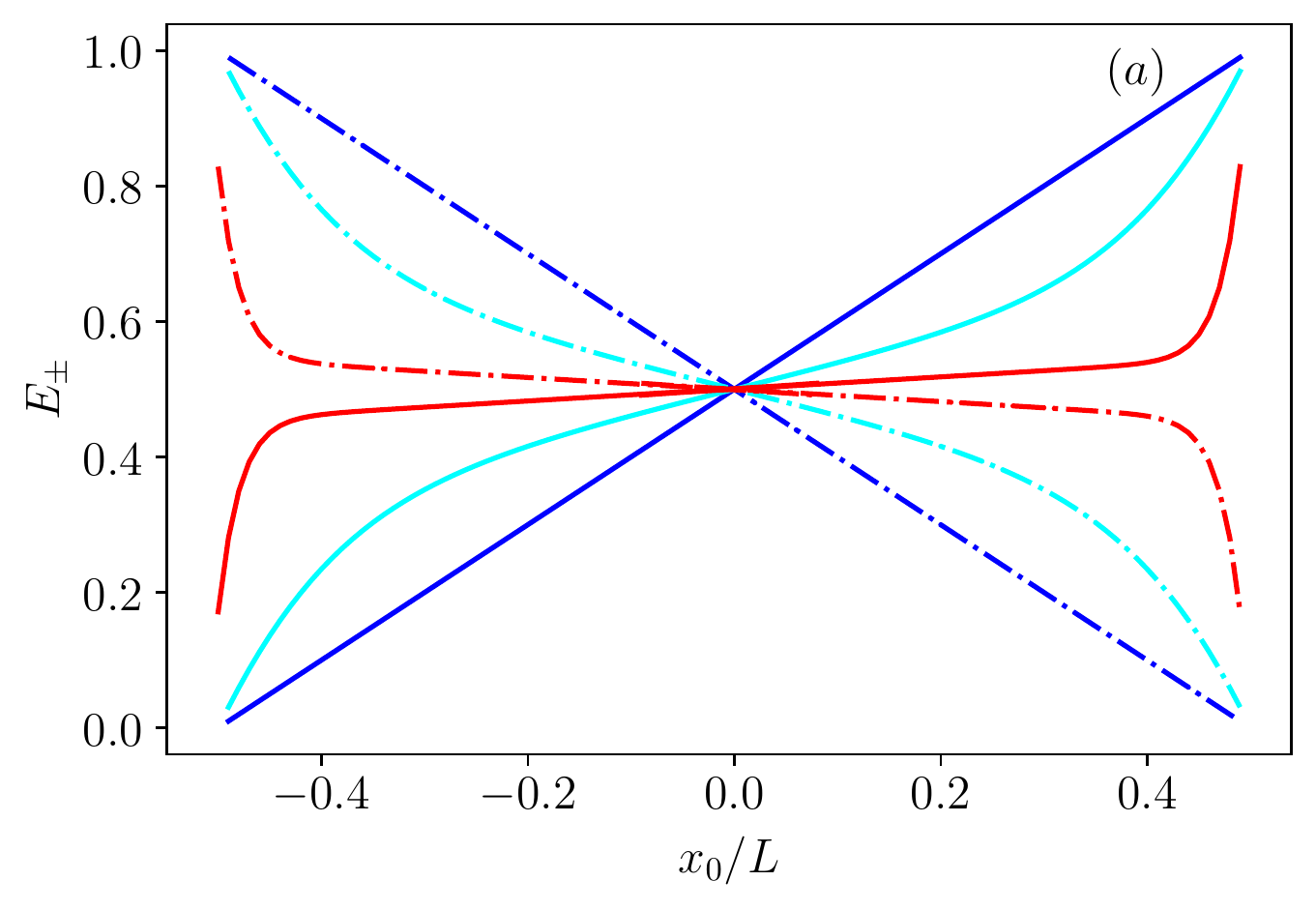}
\includegraphics[scale=0.55]{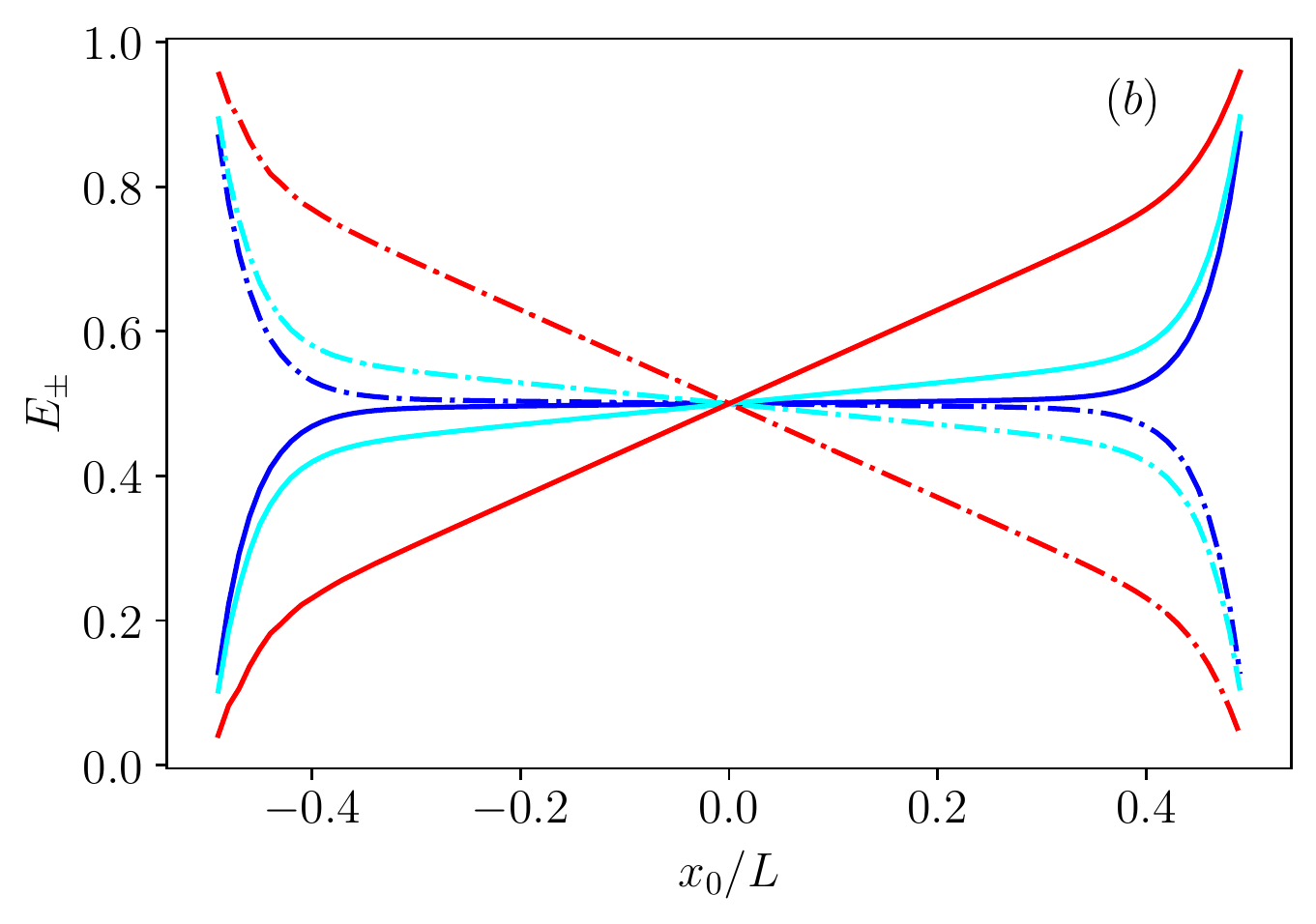}
 \caption{$E_+$ (solid curves) and $E_-$ (dot-dashed curves) as functions of $x_0/L$. Panel (a):  Fixed $\Gamma = 0.1$, $L/R=10$ and varying P\'eclet number $\text{Pe}=0$ (blue), $1$ (cyan) and $5$ (red). Panel (b):  Fixed $\text{Pe} = 3$ and  $\Gamma=0.01$ (blue), $0.1$ (cyan) and $1$ (red).}
\label{fig:act_1}
\end{figure}

Consider first a channel with a constant cross-section ($\Delta A=0$) - the  simplest  case for which, however, 
the SPs  have not been determined as yet.  
 Upon  some  straightforward algebra (see the  Suppl. Mat., Eqs.(S30) to (S42)) it is possible to derive closed-form expressions for the currents $J_\pm$ and hence, for the SPs which we depict in Fig.~\ref{fig:act_1}.
We infer from Fig. \ref{fig:act_1}(a) that upon an increase of $\text{Pe}$ the SPs $E_\pm$ become (almost) independent of the starting point, except when the  latter appears  close  to the extremities. This resembles the behavior which we observed 
for a passive particle in a channel with  $\Delta A <0$ for which the bottlenecks (entropic barriers) are at the extremities and the largest cross-section is at $x=0$. Here, the origin of such a behavior is somewhat different : For low values of  $\Gamma$ the particle does not often change the direction of its motion and travels towards the extremities of the channel  ballistically. Consequently, the larger  the propulsive force (and hence, $\text{Pe}$) is, the less sensitive are $E_{\pm}$ to the starting point. In turn, in panel (b) we plot $E_{\pm}$ as functions of $x_0/L$ with fixed $\text{Pe}$ and three different values of $\Gamma$. We realize that upon an increase of $\Gamma$ the $x_0$-dependence of the SPs approaches the linear dependence in Eq. \eqref{z3} specific  to a passive  Brownian motion in one-dimensional systems. This is, of course, not counter-intuitive - the larger $\Gamma$ is, the  more often the particle changes the direction of its motion and the dynamics becomes diffusive.

\begin{figure}[h]
\centering
\includegraphics[scale=0.55]{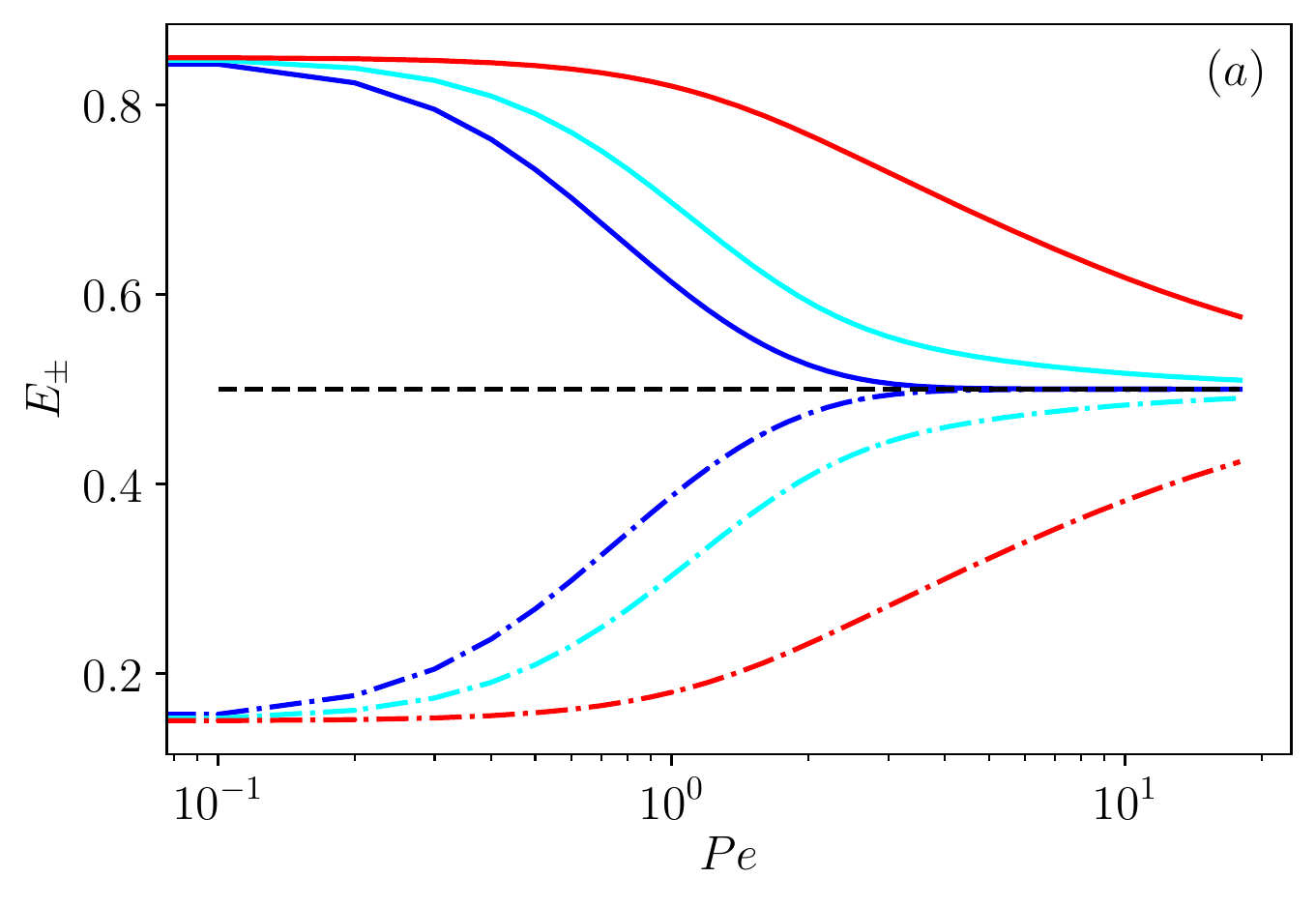}
\includegraphics[scale=0.55]{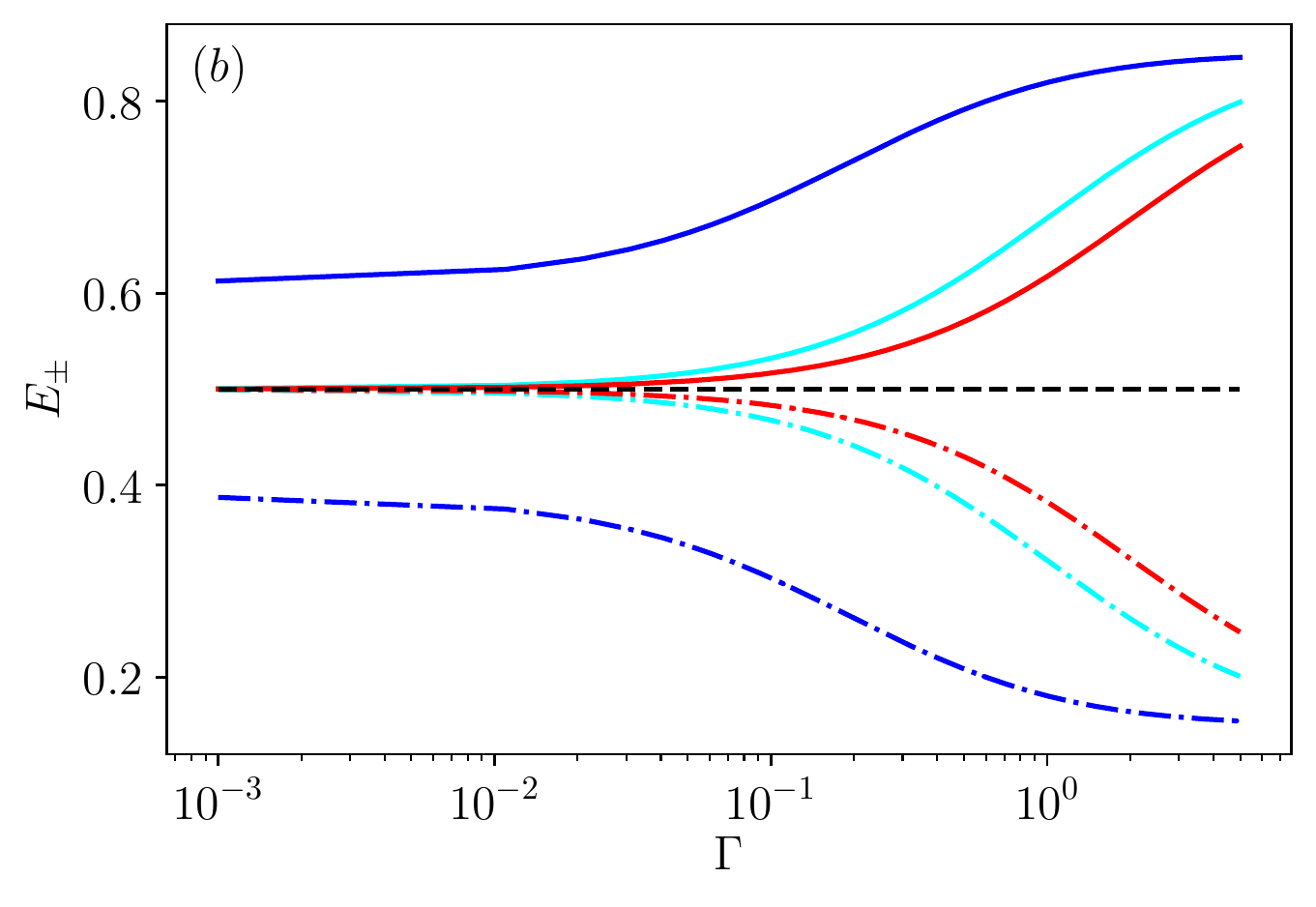}
\caption{$E_+$ (solid curves) and $E_-$ (dot-dashed curves) for $x_0/L=0.35$ 
in a channel with $L/R=10$. Horizontal dashed line defines the level $E_+ = E_-= 1/2$. Panel (a): The SPs are plotted 
as functions of the P\'eclet number for $\Gamma = 0.001$ (blue), $0.1$ (cyan) and $1$ (red). Panel (b): The SPs are plotted as functions of  $\Gamma$ for $\text{Pe}  = 1$  (blue), $5$ (cyan)  and $10$ (red). 
}
\label{fig:act2}
\end{figure}

To get an additional insight into the behavior of active particles in channels with a constant cross-section we plot in 
Fig. \ref{fig:act2} the SPs $E_{\pm}$ as functions of $\Gamma$ and $\text{Pe}$ for fixed $x_0/L = 0.35$. 
Clearly, since the starting point is close to the right extremity of the channel, one expects that $E_+ > E_-$. We observe that $E_{+}$ ($E_-$) is a monotonically decreasing (increasing)  function of $\text{Pe}$. While for small $\text{Pe}$ (for which the particle's dynamics is a passive Brownian motion) $E_+$ ($E_-$) is rather large (small) ($E_+ \approx 0.85$ and $E_- \approx 0.15$), upon an increase of $\text{Pe}$ dynamics becomes ballistic and both $E_+$  and $E_-$ tend to the same universal value $1/2$, which is rather counter-intuitive. Conversely, $E_+$ ($E_-$) is a monotonically increasing (decreasing) function of the rate $\Gamma$. Interestingly enough, in the small-$\Gamma$ limit the values of $E_+$ ($E_-$) are markedly different for small and large values of ${\text{Pe}}$ :   for  $\text{Pe} = 1$ the SP $E_+$ ($E_-$) is noticeably higher (lower) than $1/2$ (in fact, $E_- \approx 0.4$ and  $E_+ \approx 0.6$), while for $\text{Pe} = 5$ and  $10$ we have  $E_- \approx E_+ \approx 1/2$. In the limit  $\Gamma \to \infty$ the dynamics becomes diffusive and we recover the low  P\'eclet  number behavior depicted in panel (a).

\begin{figure}[h]
\centering
\includegraphics[scale=0.55]{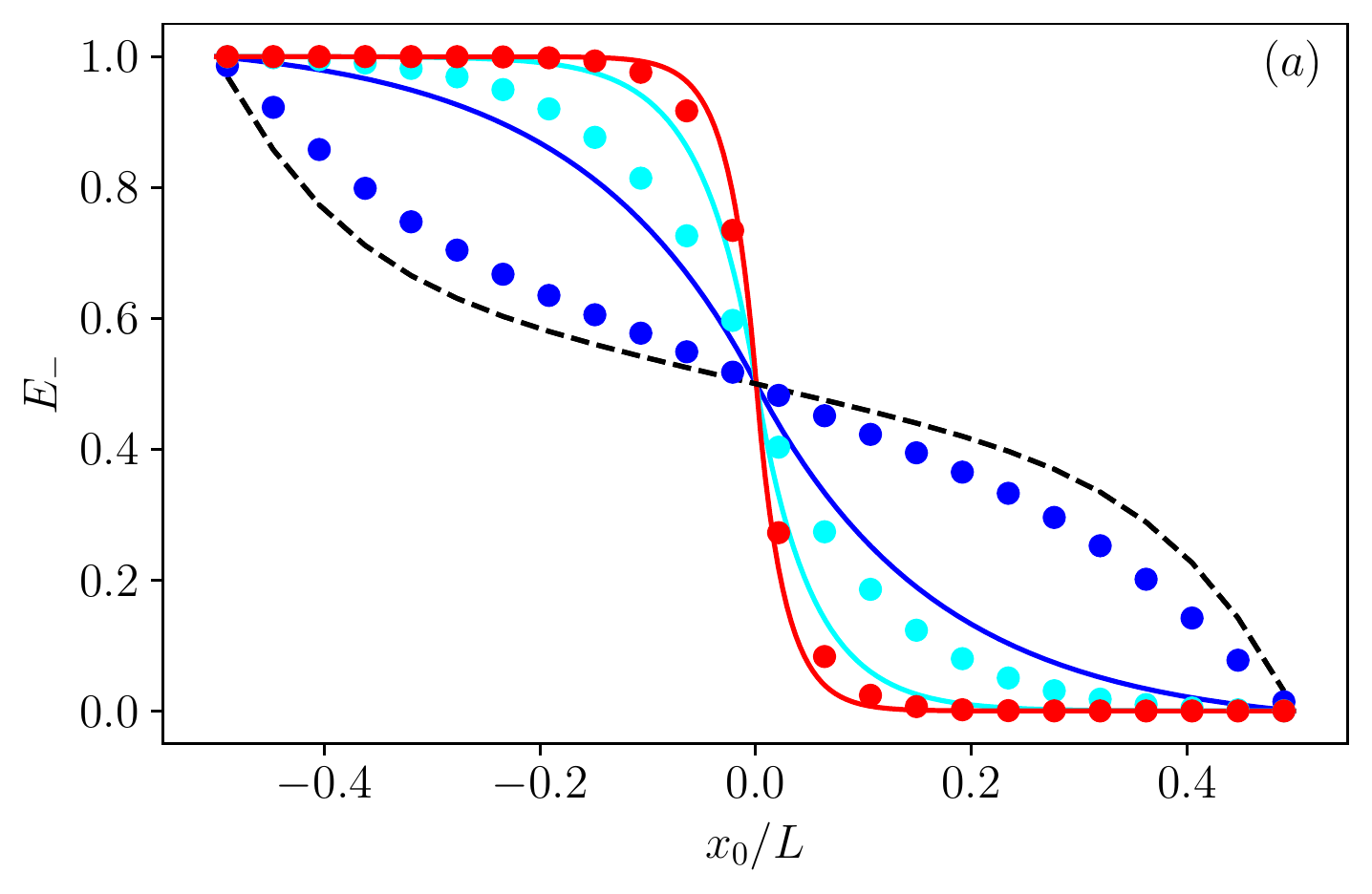}
\includegraphics[scale=0.55]{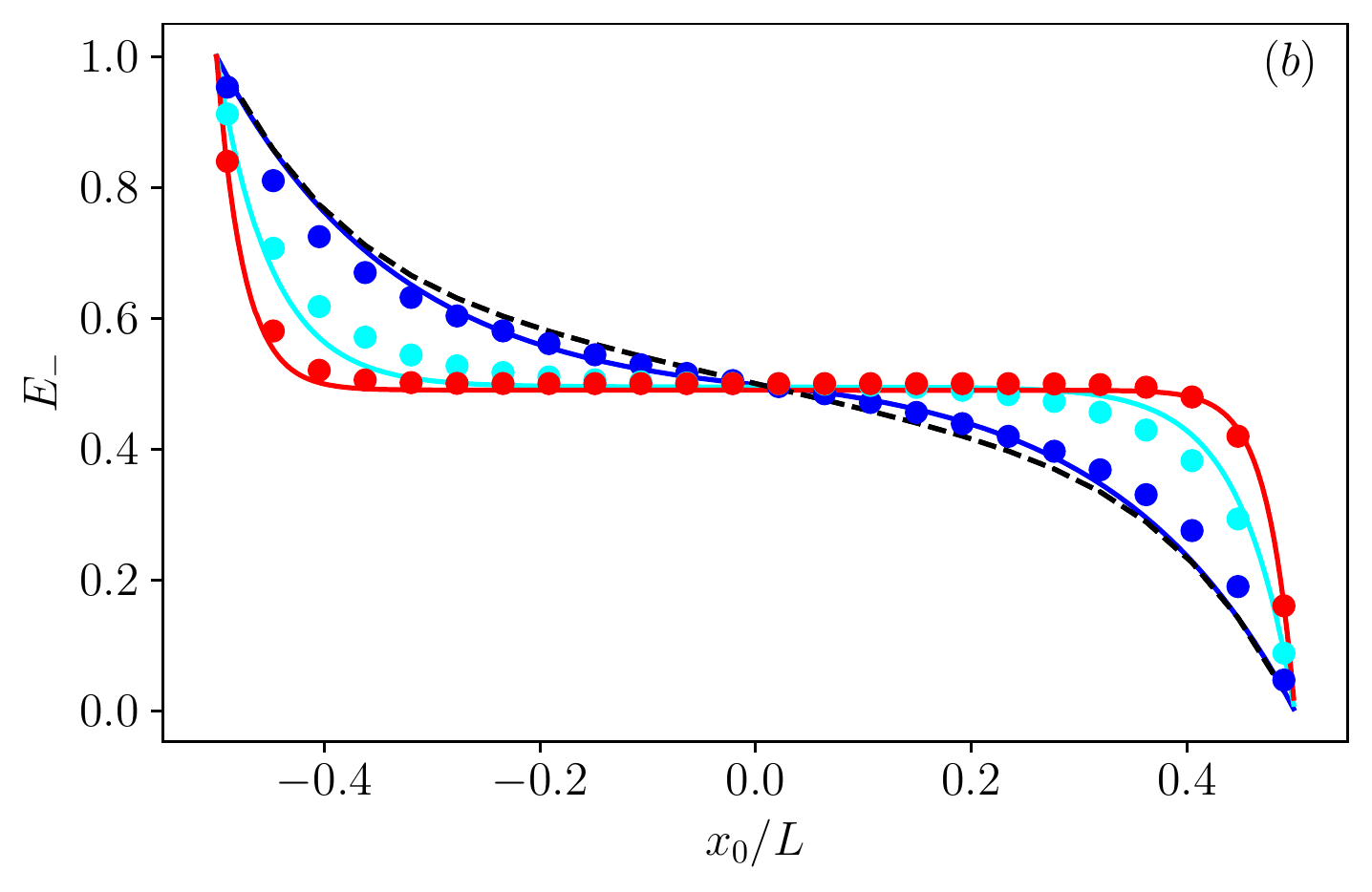}
\caption{Panel (a): $E_-$ as a function of the initial position for $\beta\Delta A/\Pe=3$ (blue), $10$ (cyan) and $20$ (red)
 and fixed $L=10R$. Solid lines indicate the respective behavior for passive particles, circles   -  for active particles with  $\Pe=1$ and $\Gamma=0.1$. Panel (b): The same for $\beta\Delta A/\text{Pe}=-3$ (blue), $-10$ (cyan) and $-20$ (red). The dashed curve depicts the analytical solution for the active particles in a channel with  a constant cross-section. }
\label{fig:act3}
\end{figure}

\begin{figure}
\centering
\includegraphics[scale=0.55]{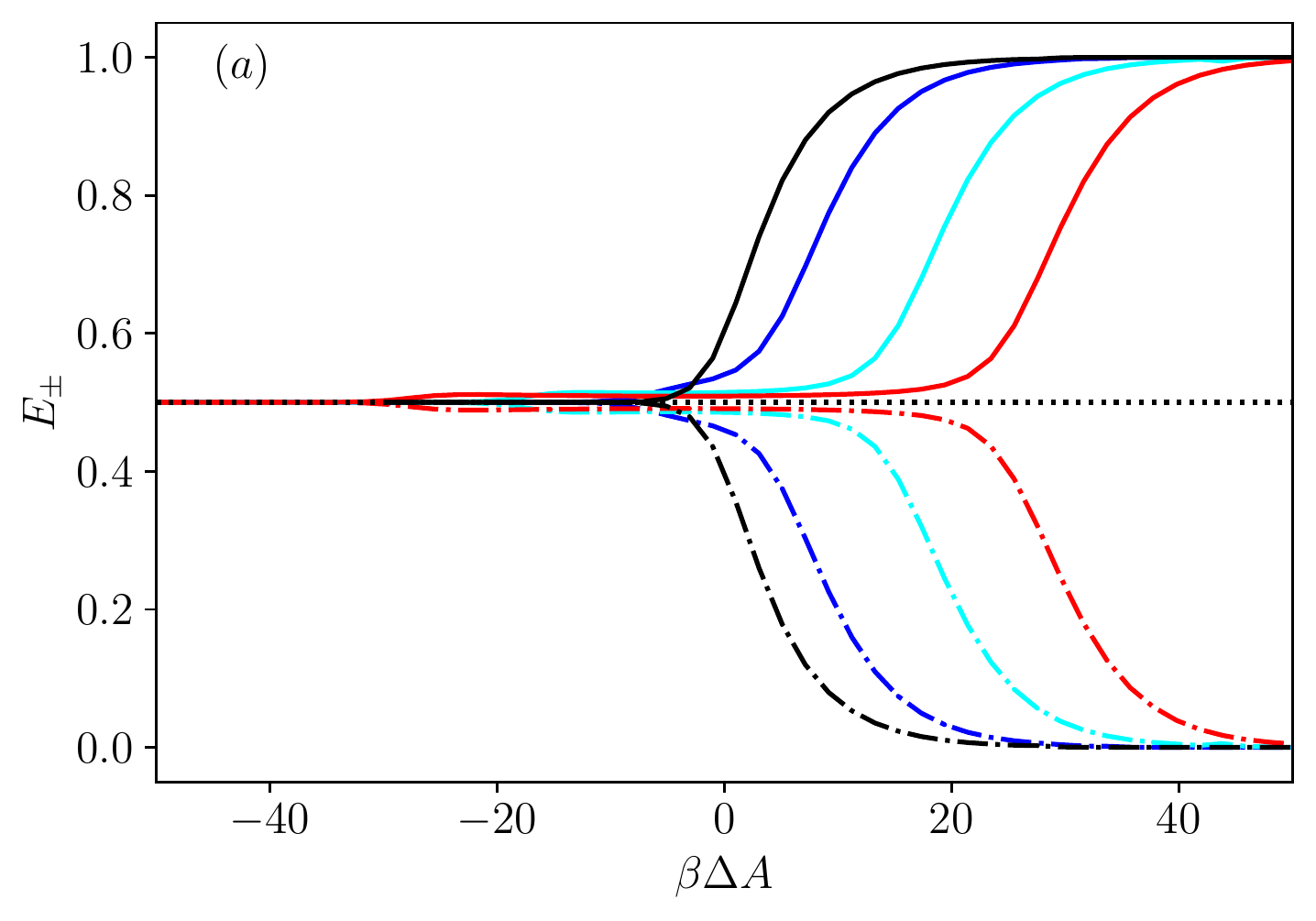}
\includegraphics[scale=0.55]{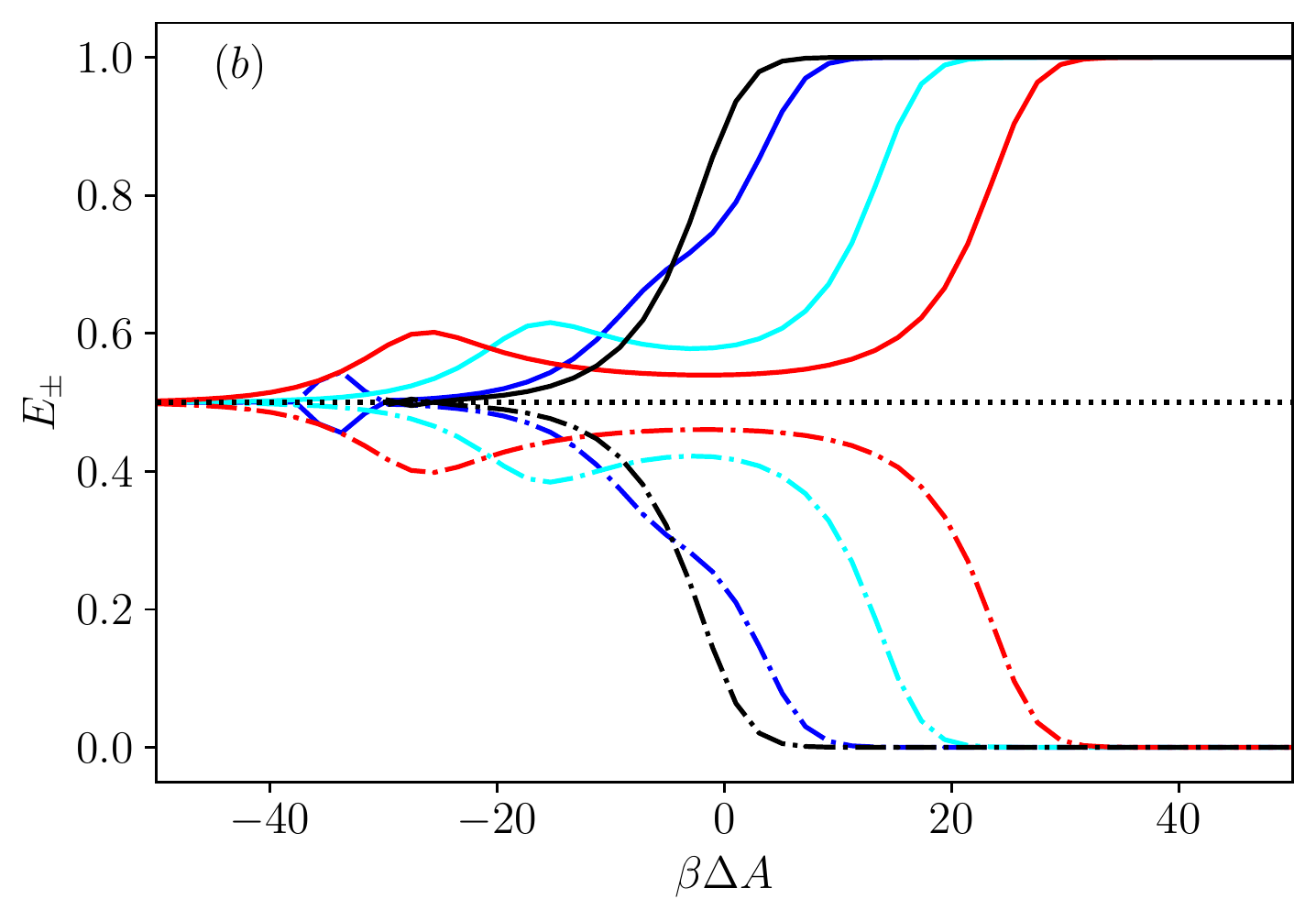}
\caption{Splitting probabilities $E_\pm$ for $x_0=0.1L$ (panel (a)) and $x_0=0.4L$ (panel (b)) as function of $\beta\Delta A$ for active particles with $\Gamma=0.1$ and $\text{Pe}=  1$ (blue), $3$ (cyan) and $5$ (red)  in a channel with $L=10R$. The black curves depict the behavior in the passive case, i.e., for $\text{Pe}=0$.
}
\label{fig:act4}
\end{figure}

 Lastly, we consider  the most difficult case  - the behavior of the SPs  for dynamics of
active particles in a channel with a varying cross-section, encoded in the effective potential $A(x)$. In this case, Eqs.\eqref{eq:smol_1} are too complicated to be solved analytically and we 
resort to a numerical analysis of these equations, which  is 
 done by using the standard scipy library in Python (see Suppl. Mat.). Our findings for the SP $E_-$ are summarized in Figs. \ref{fig:act3} and \ref{fig:act4}. 

Fig. \ref{fig:act3} displays the dependence of $E_-$ (recall that $E_+ = 1 - E_-$) on the initial position $x_0$ for fixed $\text{Pe}=0.1$, $\Gamma=3/4$ (Janus colloid case) and varying $\Delta A$. 
In this figure, circles present the results obtained numerically for active swimmers, while solid curves - an analytical solution for passive particles. Fig. \ref{fig:act3}(a) demonstrates that for positive $\Delta A$ (a bottleneck in the center), the behavior of active swimmers is very different from that of passive particles, and depends strongly on the values of both $\Delta A$ and $\text{Pe}$. 
For sufficiently large values of the parameter $\beta\Delta A/\text{Pe}$ (red circles), which limit
is realized either for large values of the barrier $\beta \Delta A$ or for small $\text{Pe}$, the SP
 $E_-$ for the active particles in channels with a varying cross-section exhibits a characteristic $S$-shaped form with a very steep dependence on $x_0$ close to the center of the channel. This implies, that once $x_0$ only slightly exceeds (or is less than) $0$, the particle is (almost) certain to reach the closest extremity without ever reaching the other one. Numerically, the value of $E_-$
 appears to be very close to the corresponding result for passive particles, which is, of course, not a counter-intuitive behavior.  In contrast,
for small $\beta\Delta A/\Pe$ (blue circles), i.e., either for large values of $\text{Pe}$ or for small values of the entropic barrier, 
$E_-$ appears to be very close to our analytical prediction obtained for active swimmers (dashed curve) moving in a constant cross-section channels which also physically quite plausible. 
Since the behavior in these limiting cases is very different, in general,  there is a strong dependence of $E_-$ on $\beta\Delta A/\Pe$ for the intermediate values of the system's parameters. We can therefore expect that particles with different activities can behave very differently in such a channel, especially if they start in the vicinity of the bottleneck.  For negative values of $\Delta A$ (entropic repulsion from the extremities), which case is presented in  Fig. \ref{fig:act4}(b), $E_-$ depends
weakly on the starting point, which resembles the behavior observed earlier for passive particles.
Further on, to highlight the difference between the passive and the active cases, in Fig. \ref{fig:act4} 
we plot $E_{\pm}$ as functions of the barrier $\Delta A$ in situations when a particle (passive or active) starts either close to the middle of the channel, at $x_0=0.1L$, or close to the right extremity of the channel, $x_0=0.4L$.  
We observe that in the active particles case the behavior of the SPs is indeed very different from that of a passive one, especially when the starting point is close to either of the extremities. While in the situation when the starting point is close to the middle of the channel (i.e., for $x_0 = 0.1 L$)
all curves look very similar with the only difference that for $\text{Pe} > 0$ they become progressively (with an increase of $\text{Pe}$) more shifted to the larger values of the barrier $\beta \Delta A$, in case when $x_0 = 0.4 L$
a remarkable non-monotonous behavior as function of $\beta \Delta A$ emerges for active particles, meaning 
that at some corrugation profiles the active particles more readily reach the right extremity. Interestingly enough, 
the position of the local maximum (minimum) of  $E_+$ ($E_-$) corresponds to $- \beta \Delta A  \simeq \text{Pe}/2$, i.e., the entropic force compensates the propulsive one. For passive particles $E_{\pm}$ are monotonously increasing
 functions of $\beta \Delta A$.

\section{Conclusion}

To conclude, we discussed here the behavior of the splitting probabilities as functions of system's parameters
for dynamics in three-dimensional axially-symmetric channels with varying cross-sections. In a standard notation, the splitting probability is the probability that either of the extremities is reached before the opposite one. In regard to the dynamical behavior, we focused on two models of random transport - standard Brownian motion and active Brownian motion.

Our analytical approach was based on a suitably generalized Fick-Jacobs approximation, which reduces an original three-dimensional model to a one-dimensional system with a spatially-varying effective potential defined as the local free energy. For standard diffusion, the latter model is exactly solvable and we derive explicit expressions for the splitting probabilities in arbitrarily shaped channels. For active Brownian motion the dynamical equations are more complicated and we find an analytical solution for constant cross-sections only. 
For more general case of a spatially-varying cross-section we resort to a numerical analysis. 

Our analysis reveals some similarities in the behavior of passive and active Brownian motions and also some distinctly different features, which can be seen as fingerprints
of the activity of particles. A more detailed discussion of the behavior in channels with a more complicated geometry
and more elaborate analytical analysis will be presented elsewhere.

\bibliography{biblio_gleb_new_1}

\onecolumngrid

\setcounter{equation}{0}
\setcounter{figure}{0}
\renewcommand\theequation{S\arabic{equation}}
\renewcommand\thefigure{S\arabic{figure}}

\section*{Appendix}

\section{Active particles}

Here we address the problem of the splitting probability
of active colloids confined to move in 1D. Indeed, the colloids can be in two possible states: moving left or moving right. Accordingly, the dynamics is controlled by the following equations: 

\begin{eqnarray}
\dot{\rho}_{\uparrow}(x,t) =  D\partial_{x}\left[\partial_{x}\rho_{\uparrow}(x,t)+\beta\rho_{\uparrow}(x,t)\partial_{x}W(x)+\beta\rho_{\uparrow}(x,t)F_{act}\right]-\alpha\rho_{\uparrow}(x,t)+\alpha\rho_{\downarrow}(x,t)\label{eq:smol_1_app}\\
\dot{\rho}_{\downarrow}(x,t) =  D\partial_{x}\left[\partial_{x}\rho_{\downarrow}(x,t)+\beta\rho_{\downarrow}(x,t)\partial_{x}W(x)-\beta\rho_{\downarrow}(x,t)F_{act}\right]+\alpha\rho_{\uparrow}(x,t)-\alpha\rho_{\downarrow}(x,t)\label{eq:smol_2_app}
\end{eqnarray}
where $F_{act}$ accounts for the active motion and with boundary conditions 
\begin{eqnarray}
\rho_{\uparrow}(x=x_0) & = & \frac{\rho_{0}}{2}\\
\rho_{\downarrow}(x=x_0) & = & \frac{\rho_{0}}{2}\\
\rho_{\uparrow}(x=\pm\frac{L}{2}) & = & 0\\
\rho_{\downarrow}(x=\pm\frac{L}{2}) & = & 0
\end{eqnarray}
In order to fulfill the above mentioned boundary conditions we split the problem into the left problem
and the right problem. Using 
\begin{align}
\rho(x,t) & =\rho_{\uparrow}(x,t)+\rho_{\downarrow}(x,t)\\
\phi(x,t) & =\rho_{\uparrow}(x,t)-\rho_{\downarrow}(x,t)
\end{align}
we get
\begin{align}
\dot{\rho}(x,t) & =D\partial_{x}\left[\partial_{x}\rho(x,t)+\beta\rho(x,t)\partial_{x}W(x)+\beta\phi(x,t)F_{act}\right]\\
\dot{\phi}(x,t) & =D\partial_{x}\left[\partial_{x}\phi(x,t)+\beta\phi(x,t)\partial_{x}W(x)+\beta\rho(x,t)F_{act}\right]-2\alpha\phi(x)
\end{align}
At steady state we get
\begin{align}
\partial_{x}\rho(x,t)+\beta\rho(x,t)\partial_{x}W(x)+\beta\phi(x,t)F_{act} & =-J/D\\
D\partial_{x}\left[\partial_{x}\phi(x,t)+\beta\phi(x,t)\partial_{x}W(x)+\beta\rho(x,t)F_{act}\right] & =2\alpha\phi(x)
\end{align}
In the case in which $\partial_{x}W(x)=0$ we get (see also EPL 134
(2), 20002): 
\begin{align}
\partial_{x}\rho(x,t) & =-J/D-\beta\phi(x,t)F_{act}\\
\partial_{x}^{2}\phi(x,t) & =\beta F_{act}J/D+\left(\left(\beta F_{act}\right)^{2}+2\alpha/D\right)\phi(x)
\end{align}
that should be solved with the boundary conditions 
\begin{eqnarray}
\rho(x=x_0) =\rho_{0}, & \rho(x=\pm\frac{L}{2}) = 0\\
\phi(x=x_0) = 0,& \phi(x=\pm\frac{L}{2}) = 0
\end{eqnarray}
The general solution of $\phi$ reads 
\begin{equation}\label{eq:gen_sol}
\phi(x)=Ae^{kx}+Be^{-kx}-\frac{\beta F_{act}J}{Dk^{2}}
\end{equation}
with 
\begin{equation}
k=\sqrt{\left(\beta F_{act}\right)^{2}+2\alpha/D}=\frac{\sqrt{\text{Pe}^2+2\Gamma}}{R}
\end{equation}
where we introduced the  P\'eclet number 
and the dimensionless hopping rate
\begin{align}
\text{Pe}=\beta F_{act} R=\frac{F_{act}}{\gamma }\frac{R}{D}=\frac{v_0 R}{D}\,,\quad \Gamma=\frac{\alpha R^2}{D}
\end{align}
and we used the Stokes-Einstein relations $\beta D=1/\gamma$ with $\gamma$ the friction coefficient of the particle.
\subsection*{Solution of the left problem}

Here we have to solve 
\begin{align}
\partial_{x}\rho_{-}(x,t) & =-J_{-}/D-\beta\phi(x,t)F_{act}\\
\partial_{x}^{2}\phi_{-}(x,t) & =\beta F_{act}J_{-}/D+\left(\left(\beta F_{act}\right)^{2}+2\alpha/D\right)\phi(x)
\end{align}
with the boundary conditions 
\begin{eqnarray}
\rho_{-}(x=x_0) =  \rho_{0}, & \rho_{-}(x=-\frac{L}{2}) =  0\\
\phi_{-}(x=x_0) = 0, &  \phi_{-}(x=-\frac{L}{2}) = 0
\end{eqnarray}
Hence we have
\begin{align}
A_{-} & =e^{-kx_0}\left[\frac{\beta F_{act}J_{-}}{Dk^{2}}-B_{-}e^{-kx_0}\right]\\
B_{-} & =\frac{\beta F_{act}J_{-}}{Dk^{2}}\dfrac{\left[1-e^{-k(x_0+L/2)}\right]}{e^{kL/2}-e^{-k(2Y+L/2)}}
\end{align}
from which we have 
\begin{align}
A_{-} & =\frac{\beta F_{act}J_{-}}{Dk^{2}}\left[e^{-kx_0}-\dfrac{1-e^{-k(x_0+L/2)}}{e^{kL/2}-e^{-k(2Y+L/2)}}e^{-2kY}\right]=\frac{\beta F_{act}J_{-}}{Dk^{2}}\mathcal{A}_{-}\\
B_{-} & =\frac{\beta F_{act}J_{-}}{Dk^{2}}\left[\dfrac{1-e^{-k(x_0+L/2)}}{e^{kL/2}-e^{-k(2Y+L/2)}}\right]=\frac{\beta F_{act}J_{-}}{Dk^{2}}\mathcal{B}_{-}
\end{align}
The general solution for $\rho$ reads 
\begin{equation}
\rho_{-}(x)=-\frac{J_{-}}{D}x-\beta F_{act}\int_{-L/2}^{x}\phi(z)dz+\Pi_{-}
\end{equation}
Substituting the formulas for $A_{-}$ and $B_{-}$ into the equation
for $\rho$ and imposing the boundary conditions we get
\begin{align}
\Pi_{-} & =-\frac{J_{-}}{D}\frac{L}{2}\\
\frac{J_{-}}{D} & = \rho_0\left[\left(\frac{\beta^2 F_{act}^2}{k^2}-1\right)\left(x_0+\frac{L}{2}\right)- \frac{\beta^2 F_{act}^2}{k^2}\left(\frac{\mathcal{A}_L}{k}\left(e^{kx_0}-e^{-kL/2} \right)+\frac{\mathcal{B}_L}{k}\left(e^{kL/2}-e^{-kx_0} \right) \right) \right]^{-1}
\label{eq:active_J_min}
\end{align}
where we used
\begin{equation}
\int_{-L/2}^{x_0}\phi(x)dx=\frac{1}{k}\left[\mathcal{A}_{-}\left(e^{kx_0}-e^{-kL/2}\right)+\mathcal{B}_{-}\left(e^{kL/2}-e^{-kx_0}\right)\right]-\frac{\beta F_{act}J}{Dk^{2}}\left(x_0+\frac{L}{2}\right)
\end{equation}

\subsection*{Solution of the right problem}

Here we have to solve 
\begin{align}
\partial_{x}\rho_{+}(x,t) & =-\frac{J_{+}}{D}-\beta\phi(x,t)F_{act}\\
\partial_{x}^{2}\phi_{+}(x,t) & =\frac{\beta F_{act}J_{+}}{D}+\left(\left(\beta F_{act}\right)^{2}+2\alpha/D\right)\phi(x)
\end{align}
with the boundary conditions 
\begin{eqnarray}
\rho_{+}(x=x_0)=\rho_0, &
\rho_{+}(x=\frac{L}{2})=0,\\
\phi_{+}(x=x_0)=0, & \phi_{+}(x=\frac{L}{2})= 0.
\end{eqnarray}
Hence we have 
\begin{align}
A_{+} & =e^{-kx_0}\left[\frac{\beta F_{act}J_{+}}{Dk^{2}}-B_{+}e^{-kx_0}\right]\\
B_{+} & =\frac{\beta F_{act}J_{+}}{Dk^{2}}\dfrac{\left[1-e^{-k(x_0-L/2)}\right]}{e^{-kL/2}-e^{-k(2Y-L/2)}}
\end{align}
from which we have 
\begin{align}
A_{+} & =\frac{\beta F_{act}J_{+}}{Dk^{2}}\left[e^{-kx_0}-\dfrac{1-e^{-k(x_0-L/2)}}{e^{-kL/2}-e^{-k(2Y-L/2)}}e^{-2kY}\right]=\frac{\beta F_{act}J_{+}}{Dk^{2}}\mathcal{A}_{+}\\
B_{+} & =\frac{\beta F_{act}J_{+}}{Dk^{2}}\left[\dfrac{1-e^{-k(x_0-L/2)}}{e^{-kL/2}-e^{-k(2Y-L/2)}}\right]=\frac{\beta F_{act}J_{+}}{Dk^{2}}\mathcal{B}_{+}
\end{align}
The general solution for $\rho$ reads 
\begin{equation}
\rho_{+}(x)=-\frac{J_{+}}{D}x+\beta F_{act}\int_{x}^{L/2}\phi_{+}(z)dz+\Pi_{+}
\end{equation}
Substituting the formulas for $A_{-}$ and $B_{-}$ into the equation
for $\rho$ and imposing the boundary conditions we get
\begin{align}
\Pi_{+} & =\frac{J_{+}}{D}\frac{L}{2}\\
\frac{J_{+}}{D} & = \rho_0\left[\left(\frac{\beta^2 F_{act}^2}{k^2}-1\right)\left(x_0-\frac{L}{2}\right)+ \frac{\beta^2 F_{act}^2}{k^2}\left(\frac{\mathcal{A}_+}{k}\left(e^{kL/2}-e^{kx_0} \right)+\frac{\mathcal{B}_+}{k}\left(e^{-kx_0}-e^{-kL/2} \right) \right) \right]^{-1}
\label{eq:active_J_plus}
\end{align}
where we used
\begin{equation}
\int_{x_0}^{L/2}\phi(x)dx=-\frac{1}{k}\left[\mathcal{A}_{+}\left(e^{kx_0}-e^{kL/2}\right)+\mathcal{B}_{+}\left(e^{-kL/2}-e^{-kx_0}\right)\right]+\frac{\beta F_{act}J}{Dk^{2}}\left(x_0-\frac{L}{2}\right)
\end{equation}
From $J_{-}$ and $J_{+}$ it is straightforward to define $\tau_\pm$ 
\begin{align}
\tau_-=\frac{\rho_0 L}{|J_-|}\\
\tau_+=\frac{\rho_0 L}{|J_+|}
\end{align}
and hence the splitting probabilities (see Eq. in the main text).

\subsection{Numerical solution}
\begin{figure}[h]
\includegraphics[scale=0.6]{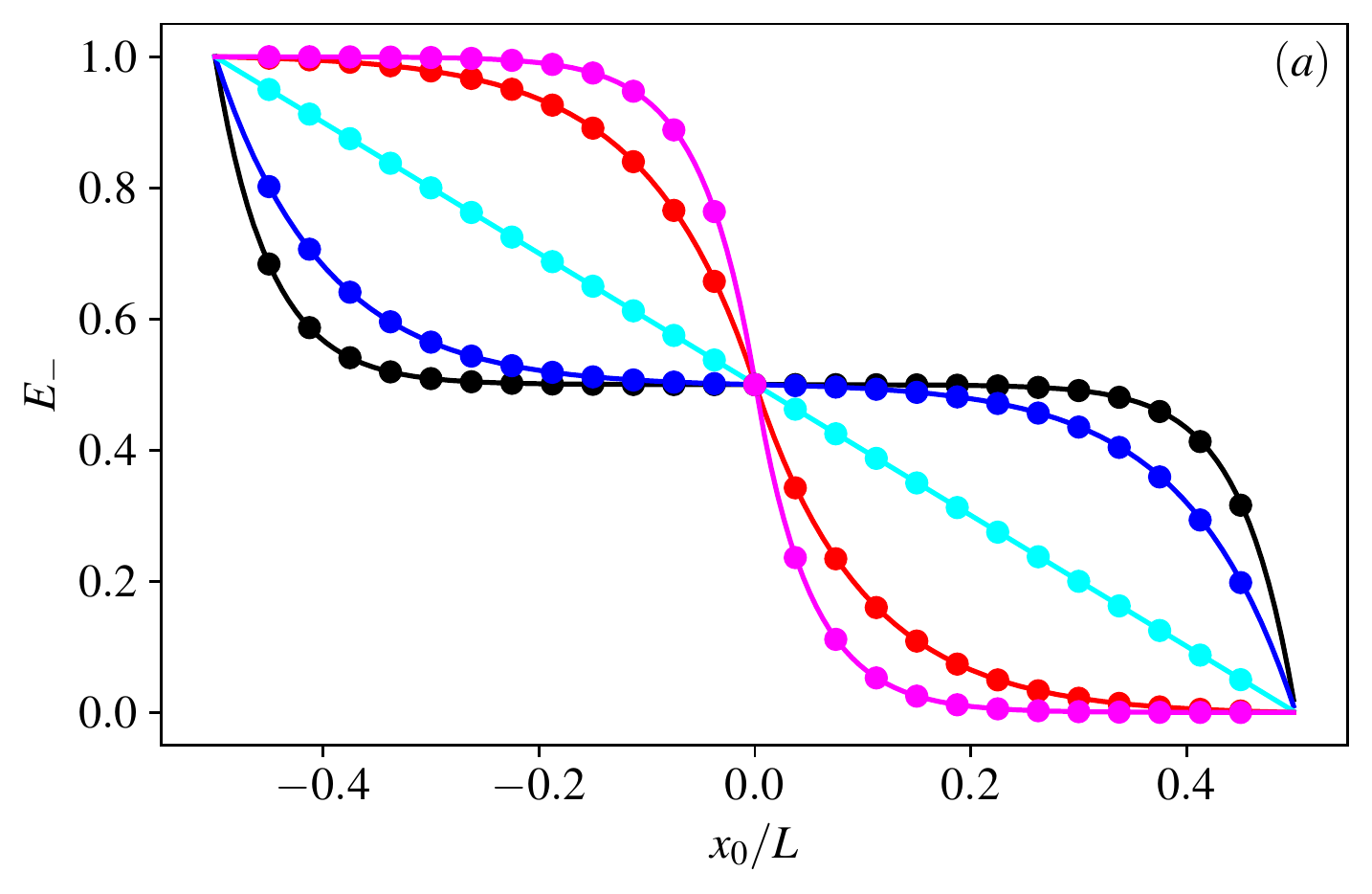}
\includegraphics[scale=0.6]{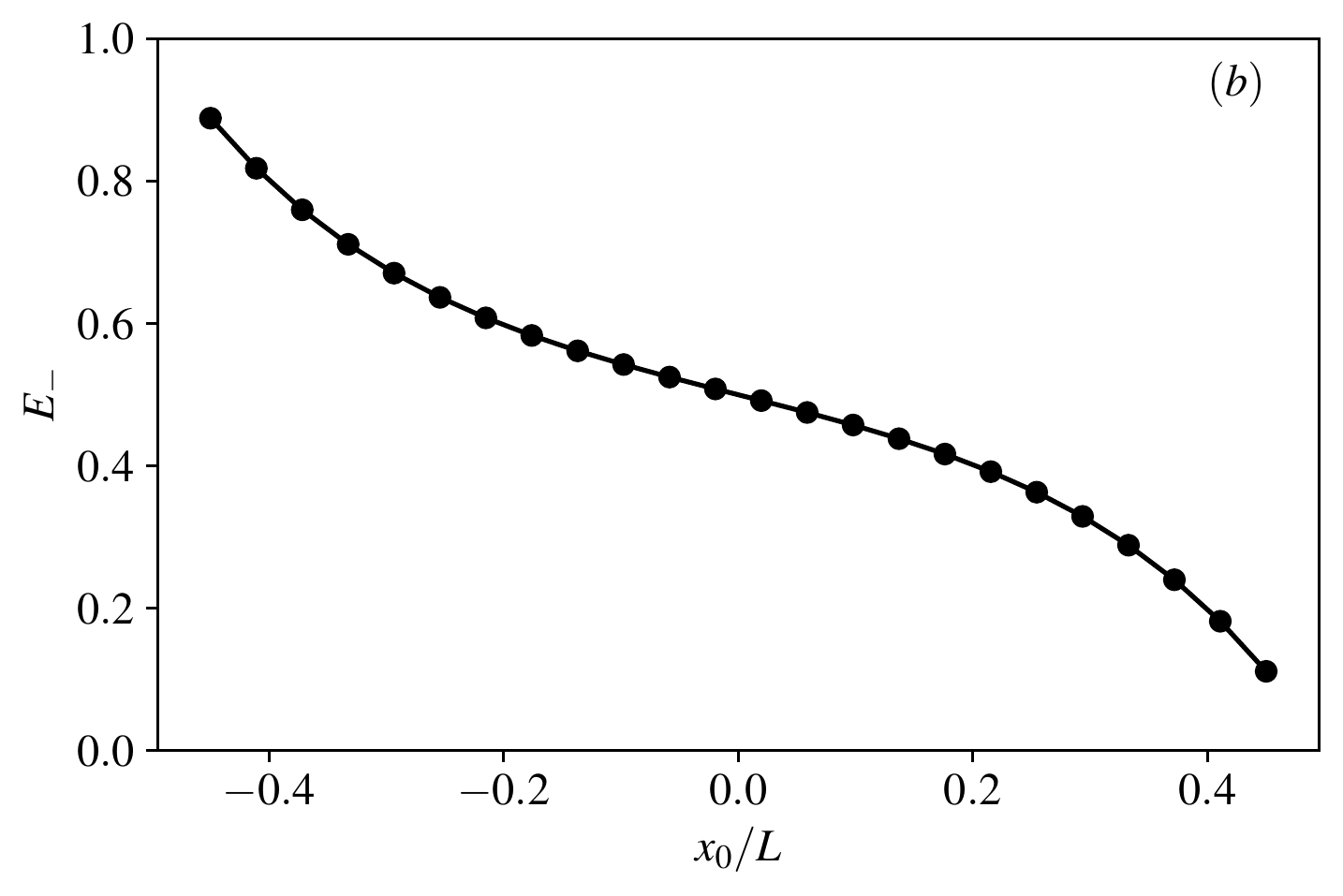}
 \caption{ Numerical (symbols) and analytical (solid curves) solution for $E_-$ as function of $x_0$ for (a) passive particles and varying $\beta\Delta A= -10,-5,5,10$ and (b) active particles with $L=10R$, $\Pe=0.5$ and $\Gamma=0.1$.}
\label{fig:act_1_app}
\label{fig:num}
\end{figure}

For arbitrary $A(x)$ Eqs.~\eqref{eq:smol_1_app},\eqref{eq:smol_2_app} can be solved numerically. To do so, we rewrite them in form of a system of first-order differential equations:
$$\mathbf{Y}'=\mathbf{M}(x)\mathbf{Y},$$
where $\mathbf{Y}=(\rho,\rho',\phi,\phi')$ and 
\begin{equation}
\mathbf{M}(x)=\left(
\begin{array}{llll}
0 & 1 & 0 & 0\\
-A''(x)& -A'(x) &-\dfrac{\Pe L}{R} &0 \\
0& 0& 0& 1\\
0& -\dfrac{\Pe L}{R} & (-2L^2/R^2)\Gamma &-A'(x) 
\end{array}
\right).
\end{equation}
With boundary conditions for the left and the right problem:
$$Y_1(\pm L/2)=0, \quad Y_3(\pm L/2)=0 \quad Y_1(x_0)=1, \quad Y_3(x_0)=0.$$ 
This system has been solved numerically using the standard Python  library \verb+scipy+. 
All calculations have been performed on a grid with $N_p=500$ nodes. The numerical solution showed good agreement with analytical solution for the case of passive particles in condfining potential and active particles in a flat channel (see Fig. \ref{fig:num}).

\end{document}